\documentclass[a4paper,12pt]{article}
 
\usepackage{bbm}
\usepackage[utf8]{inputenc}
\usepackage{placeins}
\usepackage{amssymb,amsmath,amsfonts}
\usepackage[colorlinks,linkcolor=purple,citecolor=teal]{hyperref}
\usepackage{dsfont}
\usepackage{color}
\usepackage{graphicx}
\usepackage{hyphenat}
\usepackage{wrapfig}
\usepackage{empheq}
\usepackage{textcomp}
\usepackage[caption=false]{subfig}
\usepackage{rotating}
\usepackage{wrapfig}
\usepackage{pdfpages}
\usepackage{verbatim}
\usepackage{setspace}
\usepackage{array}
\usepackage{caption}
\usepackage[pdf]{pstricks}
\usepackage{upgreek}
\usepackage{cmll}
\usepackage{latexsym}
\usepackage{braket}
\usepackage{cite}
\usepackage{epsfig}
\usepackage[left=2.4cm,top=3.3cm,right=2.4cm,bottom=3.3cm,bindingoffset=0cm]{geometry}
\usepackage[titletoc,page]{appendix}
\usepackage{multicol}
\usepackage{youngtab}

\usepackage[normalem]{ulem}     

\newcommand{\beq}{\begin{eqnarray}}
\newcommand{\eeq}{\end{eqnarray}}
\newcommand{\bea}{\begin{eqnarray}}
\newcommand{\bqa}{\begin{eqnarray}}
\newcommand{\eea}{\end{eqnarray}}
\newcommand{\be}{\begin{equation}}
\newcommand{\ee}{\end{equation}}

\newcommand{\tr}{\mathrm{tr}}

\newcommand{\calR}{\mathcal{R}}
\newcommand{\rmc}{\mathrm{c}}

\newcommand{\rmS}{\mathrm{S}}
\def\brc{\langle}
\def\ckt{\rangle}
\def\const{{\rm const}}
\def\de{\partial}

\def\nn{\nonumber}

\def\Tr{\qopname\relax o{Tr}}

\numberwithin{equation}{section}

\numberwithin{equation}{section}

\def\sp{-\!\!}

\begin{document}

\title{Dynamical Abelianization and anomalies  \\  in chiral gauge theories}

\vskip 40pt  
\author{  
Stefano Bolognesi$^{(1,2)}$, 
 Kenichi Konishi$^{(1,2)}$, Andrea Luzio$^{(3,2)}$    \\[13pt]
{\em \footnotesize
$^{(1)}$Department of Physics ``E. Fermi", University of Pisa,}\\[-5pt]
{\em \footnotesize
Largo Pontecorvo, 3, Ed. C, 56127 Pisa, Italy}\\[2pt]
{\em \footnotesize
$^{(2)}$INFN, Sezione di Pisa,    
Largo Pontecorvo, 3, Ed. C, 56127 Pisa, Italy}\\[2pt]
{\em \footnotesize
$^{(3)}$Scuola Normale Superiore,   
Piazza dei Cavalieri, 7,  56127  Pisa, Italy}\\[2pt]
\\[1pt] 
{ \footnotesize  stefano.bolognesi@unipi.it, \ \  kenichi.konishi@unipi.it,  \ \  andrea.luzio@sns.it}  
} 
\date{}

\vskip 6pt

\maketitle

\begin{abstract}

We explore the idea that in some class of strongly-coupled chiral  $SU(N)$ gauge theories 
 the infrared dynamics might be characterized by a bifermion condensate in the adjoint representation 
of the color gauge group.  As an illustration,  in this work   we revisit 
an $SU(N)$  chiral gauge theory with Weyl fermions in a symmetric  ($\psi$) and anti-antisymmetric  ($\chi$) tensor representations, together with eight
fermions in the anti-fundamental representations ($\eta$),    which we called  $\psi\chi\eta$ model in the previous investigations.
  We study the infrared dynamics of this system more carefully, by assuming dynamical Abelianization, a phenomenon  familiar from ${\cal N}=2$
supersymmetric gauge theories, and by analyzing the way various continuous and discrete symmetries are realized at low energies. 
We submit then these ideas to a more stringent test, by taking into account  
 some higher-form symmetries and the consequent mixed anomalies.
 A detailed analysis of the  mixed anomalies involving certain $0$-form $U(1)$ symmetries  and the 
color-flavor locked $1$-form ${\mathbbm Z}_N$ symmetry in the $\psi\chi\eta$ system  
 shows that the proposed infrared dynamics is consistent with it.

\end{abstract}

\newpage

\tableofcontents



\newpage

\section{Introduction}

Dynamics of strongly-coupled chiral gauge theories  is still largely unknown,  after many years of studies \cite{Raby}\sp\cite{ShiShr2}  and   in spite of their potential role in
constructing  the theory of the fundamental interactions,   
beyond  the standard  
        $SU(3)_{\rm QCD} \times (SU(2)_L\times U(1)_Y)_{\rm GWS} $
model of the strong and electroweak interactions. 

In the last few years  some renewed efforts to understand better this class of gauge theories have  been made  \cite{BKS}\sp\cite{BKL2},  mainly by using  the idea of anomaly-matching consistency requirement,  
both based on the conventional 't Hooft anomalies \cite{tHooft}, and on the recently found  generalized (e.g., 1-form)  symmetries  and mixed anomalies \cite{GKSW}\sp\cite{Anberhongson}.   
Also, the importance of the strong anomaly and its implications  has been pointed out    in the context of a large class of (generalized Bars-Yankielowicz and Georgi-Glashow) models, recently \cite{StrongAn}.

In the present work, we start to explore  earnestly  the idea that in some chiral gauge theories   bifermion condensates in the adjoint representation of the 
(strong) gauge group  form,  and play a central role in determining the infrared physics.    A possible consequence of such a condensate is dynamical Abelianization,  a phenomenon familiar  from the exact Seiberg-Witten solution of ${\cal N}=2$ supersymmetric theories \cite{SW}\sp\cite{Tachikawa}, where elementary adjoint scalar fields  are present in the theory whose vacuum expectation values (VEV)  play the crucial role in the dynamics of the theories. 
In theories of our interest  (a class of non-supersymmetric  chiral gauge theories),  such an adjoint scalar emerges as a composite field,  but nothing forbids it 
 to acquire dynamically nonvanishing vacuum expectation value (VEV),  breaking either part of color gauge symmetry,  part of the flavor symmetry, or both.
An interesting possibility is that it leads to dynamical Abelianizatoin, i.e., the gauge group is broken as
\be     SU(N)  \to     U(1)^{N-1}\;,
\ee
leaving a  weakly-coupled, IR free,  Abelian theory with a number of massless fermions.

 Such a scenario emerged  in our previous studies  \cite{BKS,BK}  as  a way of  finding a possible solution to the anomaly matching equations in the $\psi\chi\eta$  and in some other models. Let us note that in many chiral gauge theories as those studied in \cite{BKS,BK,BKL1}   even the conventional 't Hooft anomaly matching requirement  represents  a highly nontrivial constraint on the possible infrared dynamics,  in general not easy to satisfy.  

In this work  we revisit the physics of the  $\psi\chi\eta$ model more carefully,   assuming  dynamical Abelianization and reviewing the massless degrees of freedom,  consistent with  the conventional  't Hooft anomaly argument. The structure of the low-energy effective action is studied, by taking into account  all the anomalous and nonanomalous symmetries as well as the effects of the strong anomalies.  
We then examine the  generalized symmetries and  the consequent,  mixed anomalies involving some $0$-form  $U(1)$ symmetries and  $1$-form color-flavor locked center ${\mathbbm Z}_N$  symmetry.  Nontrivial anomalies found indicate that there is an obstruction in gauging simultaneously one of  the $0$-form  $U(1)$ symmetries  together with the center   ${\mathbbm Z}_N$  symmetry   (generalized 't Hooft anomalies), implying that some of the symmetries involved must be broken.   The pattern of the symmetry breaking predicted by the assumption of dynamical Abelianization  is found to fit nicely with these expectations.

\section{$\psi\chi\eta$ model and its symmetries \label{Symmetries}} 

The $\psi\chi\eta$ model  was  studied earlier  in \cite{Goity:1985tf,Eichten:1985fs,AS}   and more recently in   \cite{BKS,BK}.
  It  is  an $SU(N)$ gauge theory  with left-handed fermion matter fields  
\be   \psi^{\{ij\}}\;, \qquad  \chi_{[ij]}\;, \qquad    \eta_i^A\;,\qquad  A=1,2,\ldots 8\;,  
\ee
a symmetric tensor,  an  anti-antisymmetric tensor and eight  anti-fundamental multiplets  of $SU(N)$,  or
\be     \yng(2) \oplus  {\bar  {\yng(1,1)}}  \oplus   8  \times  {\bar  {\yng(1)}}\;.  \label{fermions2}
\ee
The model has a global $SU(8)$ symmetry. 
It  is  asymptotically free, the first coefficient of the beta function being,
\be   b_0  = \frac 13\left[ 11N-  (N+2) -(  N-2) -8  \right] =  \frac{ 9 N- 8 }{3}\;.      \label{beta0}
\ee
Such a $\beta$ function suggests that it is a very strongly coupled theory in the infrared:  it is unlikely that it flows into an infrared-fixed  CFT.  
But then some very nontrivial dynamical phenomenon must take place towards the infrared:  confinement, tumbling (dynamical gauge symmetry breaking), or something else. 
 The option that the system confines, with no global symmetry breaking and with some massless ``baryons"  saturating the 't Hooft anomalies, does not appear to be
 plausible  \cite{Goity:1985tf,Eichten:1985fs,AS},  as it would require an order $\propto N$ of the underlying fermions to form gauge-invariant baryons. The wish to understand 
what happens in the (after all,  simple)  systems such as  the $\psi\chi\eta$  model, was the driving motivation for the  renewed studies on this model \cite{BKS,BK}. Several possible 
dynamical scenarios have been found which are  all  compatible with 't Hooft's anomaly matching conditions, but the results of the analysis remained  not quite conclusive.

The system  has  three $U(1)$ symmetries,   $U(1)_{\psi}$,   $U(1)_{\chi}$, $U(1)_{\eta}$, of which two  combinations are anomaly-free. 
For convenience we will take them  below  as 
\be   \tilde U(1): \qquad   \psi\to  e^{2i\alpha} \psi\;, \quad \chi \to   e^{ -2i\alpha} \chi\;, \quad \eta \to   e^{ -i\alpha} \eta\;, \label{def:tildeU1} \ee
and
\be U(1)_{\psi\chi}: \qquad   \psi\to  e^{i  \frac{N-2}{N^*} \beta} \psi\;, \qquad \chi \to   e^{- i \frac{N+2}{N^*}\beta} \chi\;,   \qquad \eta \to   \eta \;, \label{def:U1psichi}\ee  
where 
\be N^*=GCD(N+2, N-2)\quad \text{and}\quad  \alpha,\;\beta\in (0, 2\pi)\;.\ee
Any combination of the three classical $U(1)$ symmetries which cannot be 
expressed as a linear combination of the above two, suffers from the strong anomaly.   As is well known, the consideration  
of such an anomalous symmetry also provides us with  an important information about the infrared physics.   
The famous  $U_A(1)$  problem  and its solution \cite{WittenU1}-\cite{Nath} are an example of this.  For related considerations  in the context of chiral gauge theories, see \cite{Vene,StrongAn}.
For the present model,  we will take  
\be    U(1)_{\rm an}:          \qquad   \psi \to  e^{i \delta}   \psi\;,  \qquad  \chi \to    e^{-i \delta}   \chi\;,  \qquad   \eta  \to \eta \;.   \label{anomU1} 
\ee
A nonvanishing instanton amplitude 
\be  \brc \underbrace{\psi \psi \ldots  \psi}_{N+2} \underbrace{\chi \chi \ldots  \chi}_{N-2}  \underbrace{\eta ... \eta}_{8}  \ckt \ne 0   \label{instanton}  
\ee
involving  $N+2$ $\psi$'s, $N-2$ $\chi$'s and $8$ $\eta$'s, is   indeed  not   invariant under   $U(1)_{\rm an}$ { while it is invariant under (\ref{def:tildeU1})  and   (\ref{def:U1psichi}).

There are also  anomaly-free  discrete subgroups  $({\mathbbm Z}_{N+2})_{\psi}\times ({\mathbbm Z}_{N-2})_{\chi}\times  ({\mathbbm Z}_{8})_{\eta}$   of $U(1)_{\psi} \times U(1)_{\chi} \times U(1)_{\eta}$.
Under these  ${\mathbbm Z}$'s the fields transform as 
\bea   
&&\psi  \to  e^{2\pi i  \frac{k}{N+2}} \psi\;, \qquad \ k=1,2,\ldots, N+2\;,    \label{discr1}\nn \\
&&\chi  \to   e^{ -2\pi i \frac{\ell}{N-2}} \chi\;, \qquad \ell=1,2,\ldots, N-2\;,   \label{discr2} \nn \\
&&\eta  \to    e^{ -2\pi i \frac{m}{8}} \eta\;, \qquad \ \,   m=1,2,\ldots 8\;,  \label{discr3}
\eea
which are not broken by the instantons.  However, they are not independent.
It turns out, in fact,  that $({\mathbbm Z}_{N+2})_{\psi}\times ({\mathbbm Z}_{N-2})_{\chi}\times  ({\mathbbm Z}_{8})_{\eta}$   is entirely contained inside   $SU(8) \times {\tilde U}(1)\times U(1)_{\psi\chi}$, as is easy to check.  
 The global symmetry group is connected.

Furthermore,  $\tilde U(1)\times U(1)_{\psi\chi}$ and $ ({\mathbbm Z}_{8})_{\eta}\subset  SU(8)$  has an intersection\footnote{This can be understood in a simple way. For $e^{i\alpha}\in \tilde U(1)$ and $e^{i\beta}\in U(1)_{\psi\chi}$, the composition of the two transformations acts only on $\eta$ if and only if $2\alpha+\frac{(N-2)}{N^*} \beta=2\pi \mathbbm Z$ and $-2\alpha-\frac{(N+2)}{N^*} \beta=2\pi \mathbbm Z$. Combining the two equations one obtains $\frac{8}{N^*}\alpha=2\pi \mathbbm Z$ (here we use that $\frac{(N+2)}{N^*} A - \frac{(N-2)}{N^*} B=1$ has integer solutions for $A$ and $B$, as $(N-2)/N^*$ and $(N+2)/N^*$ are co-primes). Thus $\eta\rightarrow e^{2\pi i \frac{N^*}{8}k}\eta$, which, for $k=1,\dots, 8/N^*$ forms the $\mathbbm Z_{8/N^*}$ subgroup of $(\mathbbm Z_8)_\eta \subset SU(8)$. }:
\be   \big(\tilde U(1)\times U(1)_{\psi\chi}\big) \cap  ({\mathbbm Z}_{8})_{\eta} =   {\mathbbm Z}_{8/N^*}\;. \label{Z8}
\ee
  This leads to 
the symmetry of the $\psi \chi \eta$ model: 
\be   G= \frac{ SU(N) \times  U(1)_{\psi\chi} \times  {\tilde U}(1)\times  SU(8)}{   {\mathbbm Z}_N    \times  {\mathbbm Z}_{8/ N^*} } \;.  \label{symmetry}
\ee
The division by  $ {\mathbbm Z}_N  $   is due to the fact that the color $ {\mathbbm Z}_N  $ center is shared by a subgroup of the flavor $U(1)$ groups. To see this, it is sufficient to choose  the angles 
$   \alpha =   \frac{2\pi k }{N}$,      $k \in   {\mathbbm Z}_N\;,   
$
in  (\ref{def:tildeU1});    it indeed  reduces to the
  center  ${\mathbbm Z}_{N} \subset SU(N)$   transformations of   the matter fermions,  
\be    \psi \to  e^{ 2 \cdot  \tfrac{2\pi i}{N}} \, \psi\;, \qquad  \chi \to  e^{-  2 \cdot \tfrac{2\pi i}{N}}\, \chi\;, \qquad  \eta  \to  e^{- \tfrac{2\pi i}{N}} \, \eta \;.
\label{ZN}  \ee

\section{Dynamical Abelianization
\label{fulla}}

The  aim of this work is to study the consistency of the assumption that bifermion condensates  in the adjoint representation
\be   
\langle  \psi^{ik} \chi_{kj}   \rangle  = \Lambda^3   \left(\begin{array}{ccc}
      c_1  &  &  \\     & \ddots   &   \\   &  &    c_{N}
  \end{array}\right)^i_j   \;, \qquad  \brc  \psi^{ij}  \eta_j^A  \ckt    = 0
  \;,   \label{psichicond}  
\ee
\be         c_n    \in {\mathbbm C}\;,     \qquad     \sum_n c_n =0\;, \qquad       c_m - c_n  \ne 0\;, \ \     m \ne n  \;,  \label{psichicondBis}  
\ee
(with no other particular  relations among $c_j$'s)   form  in the infrared,   inducing dynamical Abelianization of the system.

The condition of dynamical Abelianization must be made more precise. 
We require that  the condensate (\ref{psichicond}),  (\ref{psichicondBis})  induce  the symmetry breaking  
\be    
SU(N)    \to    U(1)^{N-1} \;.      \label{SUNbreaking}    \ee
As the effective composite scalar fields  $\phi \sim  \psi\chi$  are in the adjoint representation,   it is convenient to describe  them  as a linear combination, 
\be  \phi \sim  \psi\chi  =    \phi^A  T^A  =   \phi^{(\alpha)} E_{\alpha} +     \phi^{(-\alpha)} E_{-\alpha} +     \phi^{(i)} H^i \;,\label{last}  
\ee
where  $\phi^A$  are complex  fields and $T^A$ are  the Hermitian  generators of $SU(N)$ in the fundamental representation ($A= 1,2,\ldots, N^2-1$).
In  (\ref{last})  we have introduced the $SU(N)$  generators in the Cartan-Weyl basis.    $E_{\pm \alpha}$ are the  raising and lowering operators associated with positive root vectors,    $\alpha$;    $H^i$  ($i=1,2,\ldots  N-1$)  are the  generators in the Cartan subalgebra. 

    A field in the adjoint representation transforms under $SU(N)$  as
\be     \phi \to     U\, \phi  \, U^{\dagger}   \;,    \qquad     U=  e^{  i \beta^A  T^A}  \;,
\ee
i.e., as  
\be   \phi     \to      \phi +   i \beta^A [T^A,    \phi]  +\ldots\;.    \label{da1}
\ee
We recall also that the diagonal generators $T^A = H^i $ are those in the Cartan  subalgebra,  whereas the nondiagonal ones correspond to the 
pairs,
\be T^A   =   \  \frac{1}{ \sqrt{2 \alpha^2}} (E_{\alpha} +E_{-\alpha} )\;,  \          \frac{-i }{\sqrt{2 \alpha^2}}   (E_{\alpha} -  E_{-\alpha} )\;.    \label{da2}
\ee
$H^i$'s  commute with each other, and  the rest of the $SU(N)$ algebra is of the form: 
\bea   &&    [H^i, E_{\alpha}]=  \alpha_i    E_{\alpha} \;,  \qquad  
 [E_{\alpha},  E_{-\alpha}] =  \alpha   \cdot H = \sum_i \alpha_i  H^i\;,  \nonumber \\
 &&      [E_{\alpha},  E_{\beta}] =      \begin{cases}
   N_{\alpha+\beta}  E_{\alpha+ \beta}   \;,      & \text{if} \,\,   \alpha+\beta \, \, \text{is a root vector}  \;,  \\
    0  \;,    & \text{otherwise}.
\end{cases}     \label{commutators}  
 \eea
 The condition  of  dynanamical Abelianization, (\ref{SUNbreaking}),   is clearly that 
 the fields that condense are in the Cartan subalgebra, 
 \be   \phi  \sim       \psi \chi  =        \phi^{(i)} H^i \;, \qquad     \brc  \phi^i  \ckt  \ne  0\;, \qquad   \forall  i \;, \label{precise1}  
\ee
 whereas   
 \be     \brc  \phi^{(\alpha)}   \ckt =    \brc  \phi^{(-\alpha)}   \ckt =0\;,  \qquad  \forall \alpha\;. \label{precise2}  
 \ee
See below, Sec.~\ref{NGbosons}, for more about  the associated (would-be) NG bosons.

The gauge and flavor  symmetries are reduced   as:
  \beq     
  SU(N)_{\rm c} \times SU(8)_{\rm f} \times {\tilde U}(1)  \times  U(1)_{\psi\chi}  \longrightarrow  \prod_{\ell=1}^{N-1}  U(1)_{\ell}  \times  SU(8)_{\rm f} \times {\tilde U}(1)\;,
  \label{scenario3}
  \eeq
  where ${\tilde U}(1)$ is   given in (\ref{def:tildeU1}),   
    with charges
   \be       \psi:   2 \;,\qquad   \chi:   - 2\;,\qquad    \eta:   -1  \;.
  \ee
  The unbroken gauge group $ \prod_{\ell=1}^{N-1}  U(1)_{\ell}  $  is generated by the Cartan subalgebra,  
  \bea    t^1 =  \frac{1}{2} \left(\begin{array}{ccccc}1 &  &  &  &  \\ & -1 &  &  &  \\ &  & 0 &  &  \\ &  &  & \ddots &  \\ &  &  &  & 0\end{array}\right)\;, 
  \qquad  t^2 =  \frac{1}{2\sqrt{3} } \left(\begin{array}{ccccc}1 &  &  &  &  \\ & 1 &  &  &  \\ &  & -2  &   &  \\ &  &  & \ddots &  \\ &  &  &  & 0\end{array}\right)\;,
   \nn 
  \\  \qquad           \cdots,         \qquad  t^{N-1}  =  \frac{1}{\sqrt{2N(N-1)} } \left(\begin{array}{ccccc}1 &  &  &  &  \\& 1 &  &  &  \\ &  & \ddots   &   &  \\ &  &  & 1  &  \\ &  &  &  & -(N-1)\end{array}\right)\;.
  \label{gent}
  \eea
  By taking into account also the full global structure of the groups, the symmetry breaking pattern due to the (\ref{psichicond}) condensate is actually
\beq  SU(N) \times  \frac{  SU(8)_{\rm f} \times {\tilde U}(1)  \times  U(1)_{\psi\chi} }{   {\mathbbm Z}_N    \times  {\mathbbm Z}_{8/ N^*}  }    \longrightarrow    \frac{\prod_{\ell=1}^{N-1}  U(1)_{\ell}  \times SU(8)_{\rm f} \times {\tilde U}(1) }{ \prod_{\ell=1}^{N-1} {\mathbbm Z}_\ell \times {\mathbbm Z}_N \times {\mathbbm Z_2}  }  \;,
\label{fullsymbreaking}
\eeq
where
\be {\mathbbm Z}_N  = U(1)_{N-1} \cap  {\tilde U}(1) =   SU(N) \cap  {\tilde U}(1)  \;.
\ee
$ U(1)_{N-1}$ is generated by  $t^{N-1}$ in (\ref{gent})  \footnote{ By choosing   $\alpha_{\ell-1}=\alpha_{\ell}=\frac{2\pi \kappa}{\ell}$, it is easily seen that    ${\mathbbm Z}_\ell  = U(1)_{\ell} \cap  U(1)_{\ell-1}$.    Also  $\mathbbm Z_2= \tilde U(1) \cap SU(8)$.}.

The condensate  (\ref{psichicond}) leaves unbroken   a discrete   subgroup    ${\mathbbm Z}_{4/ N^*}\subset   U(1)_{\psi\chi} $:
\be     {\mathbbm Z}_{4/ N^*}   \;:  \qquad        \psi \to     e^{ i \tfrac{N-2}{N^*} \alpha} \psi\;; \quad   \chi \to     e^{ - i \tfrac{N+2}{N^*} \alpha} \chi\;,   
\ee
so that 
\be    \psi\chi  \to    e^{- i   \tfrac{4}{N^*} \alpha}    \psi\chi\;,    \label{Z4N1}
\ee
with 
\be   \alpha=    2\pi k \frac{N^*}{4}\;, \quad   k=1,2,\ldots  \frac{4}{N^*}\;.   \label{Z4N2}
\ee
Note that  ${4}/{N^*}$ is always an integer,    as  $N^*=  GCD(N+2, N-2)$  can be only one of  $1,2,4$.        We however note that 
 $\mathbbm Z_{4/N^*}$ is a subgroup of  $SU(8)\times \tilde U(1)$ \footnote{To see this, note first  that  $SU(8)\times {\tilde U}(1)$ contains a discrete subgroup   ${\mathbbm Z}_4$ acting on $\psi$ and $\chi$ by phases $\pm 2\pi k /4$, $k=1,\ldots, 4.$     Depending on $N$,   $\mathbbm Z_{4/N^*}$ of  (\ref{Z4N1}) can be seen to be  ${\mathbf 1}$,   ${\mathbbm Z}_2$ or  ${\mathbbm Z}_4$, always  in     
 ${\mathbbm Z}_4\subset  SU(8)\times {\tilde U}(1).$ 
 }.    Thus  the global unbroken symmetry group in (\ref{fullsymbreaking}) is still connected.

Another discrete symmetry which remains unbroken by the condensate  is   
 $ {\mathbbm Z}_{  N^*}  $,
\bea   && {\mathbbm Z}_{N^*} \;:         \qquad     \psi\to  e^{2\pi i  \frac{p}{N^*}} \psi\;, \quad 
 \chi \to   
e^{ -2\pi i\frac{p}{N^*}} \chi\;,   \label{znspsichi}  
  \eea
($p =  1, \ldots, { N}^*$). 
This is a subgroup \footnote{Just take   $k =p \frac{N+2}{N^* }$ and  $\ell =p \frac{N-2}{N^* }$   in  (\ref{discr3}). }  of  the nonanomalous, discrete  $({\mathbbm Z}_{N+2})_{\psi} \times ({\mathbbm Z}_{N-2})_{\chi}$ symmetries, both of which are broken by the condensate.
$ {\mathbbm Z}_{  N^*}  $   is   also a  subgroup of nonanomalous, unbroken continuous   $SU(8)\times \tilde U(1)$
 \footnote{
This can be seen  by taking   $e^{i\frac{2\pi p}{2N^*}} \in   \tilde U(1)$    
and $e^{2\pi i \frac{p}{2N^*}}    \in    \mathbbm  Z_8   \subset   SU(8)$.}.

 The pattern of the gauge symmetry breaking is somewhat reminiscent of  what happens in the ${\cal N}=2$ supersymmetric gauge theories. Indeed the massive spectrum  will contain 't Hooft-Polyakov magnetic monopoles, as well as the massive $SU(N)/U(1)^{N-1}$  gauge bosons. 
     Note however that these monopoles are not in a semiclassical regime. The coupling constant at the scale of symmetry breaking is not small but of order one, $g^2 \sim1$. Thus the monopole size and its Compton length are comparable; it is a soliton in a  highly quantum regime. 
 Our system is analogous to the   ${\cal N}=2$ supersymmetric gauge theories in the so-called  quark vacua, where the bare quark mass is cancelled by the adjoint field VEV.  
In the absence of the moduli space of vacua here  it is reasonable to assume that our low-energy system describes the photons of the electric  $\prod_{\ell=1}^{N-1}  U(1)_{\ell}  $
 theory.  Our system is analogous to the  ${\cal N}=2$  Seiberg-Witten theories
 outside the so-called marginal stability curves  \footnote{Due to the phenomenon of isomonodromy 
  the spectrum of the stable particles of the system changes when crossing some subspace of  the vacuum moduli space.    The phenomenon has been studied in detail for $SU(2)$ Seiberg-Witten theory \cite{Ferrari}\sp\cite{Ritz}. }, though perhaps not far from one.

\subsection{'t Hooft anomaly matching   \label{matching} }
  
The fields  $\eta^A_i$ which do not participate in the condensate  remain  massless and  weakly coupled  to the gauge bosons from the Cartan subalgebra which we will refer to as the photons,   in the infrared. Also, some of the fermions $\psi^{ij}$  do not participate in the condensates. Due to the fact that 
$\psi^{\{ij\}}$ are symmetric whereas $\chi_{[ij]}$ are antisymmetric,  actually only the nondiagonal elements of   $\psi^{\{ij\}}$  condense and get mass. The diagonal 
fields $\psi^{\{ii\}}$, $i=1,2,\ldots, N$, together with  all of  $\eta^a_i$  remain massless.   Also there is one physical  NG boson (see Sec~\ref{NGbosons}  below). 
\begin{table}[h!t]
  \centering 
  \begin{tabular}{|c|c|c |c| }
\hline
$ \phantom{{{   {  {\yng(1)}}}}}\!  \! \! \! \! \!\!\!$   & fields      &  $ SU(8)$     &   $ {\tilde U}(1)   $  \\
 \hline
  \phantom{\huge i}$ \! \!\!\!\!$  {\rm UV}& $\psi$      &    $   \frac{N(N+1)}{2} \cdot  (\cdot) $    & $  \frac{N(N+1)}{2} \cdot (2)$    \\
  &  $\chi$      &    $   \frac{N(N-1)}{2} \cdot  (\cdot) $    & $  \frac{N(N-1)}{2} \cdot (-2)$        \\
 &$ \eta^{A}$      &   $ N \, \cdot  \, {\yng(1)}  $     &   $  8N \cdot  (-1) $ \\
   \hline 
  \phantom{\huge i}$ \! \!\!\!\!$  {\rm IR}&       $ \psi^{ii}  $      &  $ N \cdot ( \cdot)   $        &    $  N \cdot (2) $   \\
     &  $ \eta^{A} $      &  $ N \, \cdot  \, {\yng(1)}  $        &    $  8  N \cdot (-1) $   \\
\hline
\end{tabular}  
  \caption{\footnotesize  Full dynamical Abelianization in the $\psi\chi\eta$ model.}
\label{Simplest}
\end{table}
All of the anomaly triangles,    $[SU(8)]^3$,    $  {\tilde U}(1)-[SU(8)]^2$,   $[{\tilde U}(1)]^3$,      ${\tilde U}(1)-[{\rm gravity}]^2$,  $  {\mathbbm Z}_{  N^*}-[SU(8)]^2$,       $ {\mathbbm Z}_{  N^*}-[{\rm gravity}]^2$
are  easily seen to match, on inspection of  Table~\ref{Simplest}.  Perhaps the only nontrivial ones are the ones that do not involve $SU(8)$.  For ${\tilde U}(1)-[{\rm gravity}]^2$ we have  
\be    \underbrace{ 2 \cdot  \frac{N(N+1)}{2}  -   2 \cdot    \frac{N(N-1)}{2} -  8 \cdot N }_{\rm UV}= 
  \underbrace{2 \cdot   N  -     8 N }_{\rm IR} =   - 6N   \;, 
  \ee
for $[{\tilde U}(1)]^3$ we have  
\be    \underbrace{ 8 \cdot  \frac{N(N+1)}{2}  -   8 \cdot    \frac{N(N-1)}{2} -  8 \cdot N }_{\rm UV}= 
  \underbrace{8 \cdot   N  -     8 N }_{\rm IR} =  0   \;,
  \ee
for   $ {\mathbbm Z}_{  N^*}-[{\rm gravity}]^2$ we have  
\be    \underbrace{ 1 \cdot  \frac{N(N+1)}{2}  -   1 \cdot    \frac{N(N-1)}{2}   }_{\rm UV}= 
  \underbrace{1 \cdot   N   }_{\rm IR} =   N   \;. 
  \ee
\begin{table}[h!t]
  \centering 
  \begin{tabular}{|c|c|c|c|c|c|c|c|c| }
\hline
$ \phantom{{{   {  {\yng(1)}}}}} \!  \! \! \! \! \!\!\!$   & fields    & $U(1)_1$ & $U(1)_2$  &   $\cdots $  &  $U(1)_{N-1}$    &  $ SU(8)$     &   $ {\tilde U}(1)   $    &   $ U(1)_{\rm an} $      \\
 \hline
   $\psi$  $ \phantom{{{   {  {\yng(1)}}}}}\!  \! \! \! \! \!\!\!$   & $\psi^{11}$   &   $1$     &   $\frac{1}{\sqrt{3}}$        &  $\cdots$   &  $\frac{2}{\sqrt{2N(N-1)}}$    &    $  (\cdot) $       &    $ 2$     & $ 1$      \\
                                                              &  $\psi^{22}$  &   $-1$   &   $\frac{1}{\sqrt{3}}$       &  $\cdots$  &   $\frac{2}{\sqrt{2N(N-1)}}$    &     $  (\cdot) $       &   $ 2$   & $ 1$   \\
                                                              &  $\psi^{33}$  &   $0$    &   $-\frac{2}{\sqrt{3}}$   &  $\cdots$  &   $\frac{2}{\sqrt{2N(N-1)}}$   &     $  (\cdot) $       &   $ 2$   & $ 1$   \\
                                                              &  $\vdots$       &             &                   &                  &      $\vdots$      &       $\vdots$       &    $\vdots$     & $\vdots$     \\
                                                              &  $\psi^{NN}$  &   $0$   &   $0$          &  $\cdots$      &   $-  \frac{2(N-1)}{\sqrt{2N(N-1)}}$     &     $  (\cdot) $       &          $ 2$    & $ 1$       \\
    \hline 
  $\eta$  $ \phantom{{{   {  \bar{\yng(1)}}}}}\!  \! \! \! \! \!\!\!$ &   $ \eta^a_1$  &    $ -\frac{1}{2}$       &    $-\frac{1}{2\sqrt{3}}$          &  $\cdots$  &      $-\frac{1}{\sqrt{2N(N-1)}}$    &       $ \yng(1) $   &    $  -1  $      & $0$    \\
                                 &   $ \eta^a_2$  &  $\frac{1}{2}$       &    $-\frac{1}{2\sqrt{3}}$          &  $\cdots$  &      $-\frac{1}{\sqrt{2N(N-1)}}$   &       $ \yng(1) $   &    $  -1  $    & $0$   \\
                                 &   $ \eta^a_3$  &  $0$                      &    $\frac{1}{\sqrt{3}} $          &  $\cdots$  &     $-\frac{1}{\sqrt{2N(N-1)}}$   &       $ \yng(1) $   &    $  -1  $    & $0$     \\ 
                                                     &  $\vdots$         &           &                   &                &                    &       $\vdots$     &  $\vdots$     &  $\vdots$       \\
                                                             &   $ \eta^a_N$  &   $ 0 $      &    $0$          &  $\cdots$  &     $\frac{N-1}{\sqrt{2N(N-1)}}$   &       $ \yng(1) $   &    $  -1  $     &    $0$     \\
                                                                 \hline 
   $\pi$  $ \phantom{{{   {  {\yng(1)}}}}}\!  \! \! \! \! \!\!\!$  &   ${\tilde \phi} \sim  (\psi\chi)^1_1$  & $0$  &    $0$      &    $\cdots$   &   $0$             &      $ (\cdot) $      &    $  0  $  &  $0$    \\
\hline
\end{tabular}  
  \caption{\footnotesize Massless fermions in the infrared in the $\psi\chi\eta$ model and their charges with respect to the unbroken symmetry groups.  The massless NG boson
  carries  no charges with respect to the unbroken symmetries, see Sec.~\ref{NGbosons}.   The two  (nonanomalous and anomalous)   $U(1)$ symmetries
  $ {\tilde U}(1)   $    and    $ U(1)_{\rm an}   $
   which are not affected by the    $\psi\chi$  condensate are defined in  (\ref{anomU1}).    $u(1)^{N-1} \subset su(N)$ are taken in Cartan subalgebra, satisfying the orthogonality relations, $\Tr (T^a T^b) \propto  \delta^{ab}$.        
 }
\label{Simplest2}
\end{table}
The massless fermions  in the infrared are shown again in Table~\ref{Simplest2},   where their quantum numbers with respect to the weak 
 $\prod_{\ell=1}^{N-1}  U(1)_{\ell}  \subset SU(N)$ are also shown.

\subsection{Nambu-Goldstone (NG) bosons   \label{NGbosons}  }

Discussion of  the infrared physics  implied by the dynamical Abelianization  requires also understanding of  the massless bosonic degrees of freedom,  besides the fermions in 
Table~\ref{Simplest2}.  As will be seen below, there is one physical massless $U(1)$  NG boson in this system.

Before discussing the $U(1)$  NG boson, however,  let us briefly comment on the     $SU(N)$ gauge symmetry breaking, discussed in  Sec.~\ref{fulla}.  
The diagonal generators in Cartan subalgebra $H^i$   correspond to the unbroken   $U(1)^{N-1}$  Abelian group, whereas   the broken generators of $SU(N)/U(1)^{N-1}$
are  the nondiagonal $E_{\alpha}$ and  $E_{-\alpha}$'s.     Recalling the commutation relations   (\ref{commutators}),   it is easy to see the   NG boson (interpolating) fields ${\tilde \phi}$
which get transformed into  the ones which condense,  see (\ref{WTI1})-(\ref{WTI2}),  are   $\phi^{(\alpha)}$,  $\phi^{(-\alpha)}$  of   (\ref{last}).   They are just the nondiagonal
$N(N-1)/2 \cdot 2 = N^2-N$  components of $(\psi\chi)^i_j$ composite scalars.  The would-be NG bosons $\pi^{\alpha}$    associated with the  currents $J_{\mu}^{\alpha}$, $J_{\mu}^{-\alpha}$,
\be     \brc 0 |  J_{\mu}^{\alpha}  | \pi^{\beta} \ckt  =  F_{\pi}  \delta^{\alpha \beta}\;,  \qquad   \brc  \pi^{\beta} | \phi^{-\alpha}(0) | 0\ckt   \propto  \delta^{\alpha \beta}\;,    
\ee
are eaten by the  Englert-Brout-Higgs mechanism making the $SU(N)/U(1)^{N-1}$  gauge bosons massive.

Let us now focus our attention on the $U(1)$  NG boson.  
As noted already, there are three $U(1)$ symmetries in the model, the two nonanomalous ones  ${\tilde U}(1)$  and  $U(1)_{\psi\chi}$ in (\ref{def:tildeU1}), (\ref{def:U1psichi})
  and  an anomalous $U(1)_{\rm an}$ in  (\ref{anomU1}). 
The associated currents are
\bea
  &&J_{\psi\chi}^{\mu} =   i     \left\{     \tfrac{N-2 }{N^* }   \,   {\bar \psi}  {\bar \sigma}^{\mu}   \psi -        \tfrac{N+2 }{N^* } \, {\bar \chi}  {\bar \sigma}^{\mu}   \chi  \right\} \;,
   \qquad  \de_{\mu}   J_{\psi\chi}^{\mu}  = 0\;,    \label{current1}   \\
  &&  \ {\tilde J}^{\mu}   =     i  \,   \Big\{    2  \, {\bar \psi}  {\bar \sigma}^{\mu}   \psi -     2  \, {\bar \chi}  {\bar \sigma}^{\mu}   \chi   -      {\bar \eta^a}  {\bar \sigma}^{\mu}   \eta^a  \Big\}  \;,
   \qquad \  \   \   \de_{\mu}    {\tilde J}^{\mu}   = 0\;,  \phantom{\frac{1}{2}}   \label{current2}\\
  && \, J_{\rm an}^{\mu}  =      i  {\bar \psi}  {\bar \sigma}^{\mu}   \psi -    i  {\bar \chi}  {\bar \sigma}^{\mu}   \chi   \;,   \qquad  \qquad  \qquad  \qquad  \   \ \  \,  \de_{\mu}   \,  J_{\rm an}^{\mu}  =     \frac{ 2 g^2}{32 \pi^2}   G_{\mu \nu}  {\tilde G}^{\mu \nu}  \;, 
  \label{current3}
\eea
and the associated charges are\footnote{Note that  in the two-component spinor notation of  Wess and Bagger,   $ {\bar \sigma}^{0} = - i $\;, and    $ {\bar \psi}  \equiv   \psi^{\dagger}$.}
\bea
  &&Q_{\psi\chi} =  \int   d^3 x  \left(     \tfrac{N-2 }{N^* }   \,     {\bar \psi}  \psi -          \tfrac{N+2 }{N^* }  {\bar \chi}  \chi  \right) \;,  \nonumber \\
  &&  \ \ \ {\tilde Q}=    \int   d^3 x      \,   \left(    2  \, {\bar \psi}   \psi -     2  \, {\bar \chi}   \chi   -      {\bar \eta^a}     \eta^a  \right)  \;,    \nonumber \\
  && Q_{\rm an} =    \int   d^3 x     \, ( {\bar \psi}  \psi   -   {\bar \chi}  \chi  )    \;.
\eea
It follows from the standard quantization rule that    ($ (\psi\chi)^n_m \equiv   \psi^{nk} \chi_{k m}$)
\bea
 &&   [Q_{\psi\chi},  (\psi\chi)^n_m ] =           \frac{4}{N^*}      (\psi\chi)^n_m \;, \label{Commutator1}   \nn  \\
   && \ \  \  [{\tilde Q},  (\psi\chi)^n_m ] =    (2 - 2  )   (\psi\chi)^n_m   =  0  \;,\label{Commutator2} \phantom{\frac{1}{2}} \nn  \\
   &&    \,  [Q_{\rm an} ,   (\psi\chi)^n_m ]  =   0 \;.   \phantom{\frac{1}{2}} \label{Commutator3}  
\eea
It is seen  from these  that the condensates (\ref{psichicond}), (\ref{psichicondBis})  (i.e.,    the diagonal    $\brc (\psi\chi)^n_n \ckt$)  
break   spontaneously  only  the   nonanomalous $U(1)_{\psi\chi}$   symmetry. 
Both     ${\tilde U}(1)$  and   
$U(1)_{\rm an}$  remain  unbroken. 
There is only one massless (physical)  NG boson  in the system.

Also,  it is seen easily that among  the diagonal $(\psi\chi)^n_n$'s   there is only one {\it  independent}  component,  which can be taken e.g., 
\be   U(x) =    (\psi\chi)^1_1(x) \;,  
\ee
that  is transformed into a field which acquires a nonvanishing 
VEV.\footnote{This can be seen by considering the linear combinations  such as $c_2   (\psi\chi)^1_1-  c_1   (\psi\chi)^2_2$,  $c_3   (\psi\chi)^2_2 -   c_2   (\psi\chi)^3_3$, etc., whose VEV's all vanish.  
     }    A simple chiral Ward-Takahashi identity (see Appendix~\ref{WT})  then shows   that    $J_{\psi\chi}^{\mu} $  generates from the vacuum the massless boson,  
which can be   described by the  interpolating field,    
$ U(x)  =  (\psi\chi)^1_1(x)$.

It is instructive  to compare  the situation here with the fate of the $U(1)$ symmetries in  QCD.    In QCD, the bifermion condensate is of the form,  ${\bar \psi_R} \psi_L$.   
It is invariant under $U(1)_V$ and noninvariant under  $U(1)_A$.  By appropriately choosing the phases of $\psi_L$ and $\psi_R$  the condensate
$\brc  {\bar \psi_R} \psi_L\ckt$ can be chosen to be real;  the (would-be) NG boson of the  broken  $U(1)_A$  symmetry then corresponds to the imaginary part of   
${\bar \psi_R} \psi_L$.   Due to the effects of the strong anomaly, this would-be NG boson gets mass.

Here  the condensate  $\brc \psi\chi \ckt$   is invariant under the ${\tilde U}(1)$ as well as under the anomalous  $U(1)_{\rm an}$.     Only the nonanomalous 
   $U(1)_{\psi\chi}$   is  broken by the condensate:    the quantum fluctuations of   $(\psi\chi)^1_1(x)$ contain  one  physical, massless  NG boson, $\pi$.

\subsection{The low-energy effective action   \label{LE}}   

The massless degrees of freedom in the infrared are thus   the gauge bosons  $A_{\mu}^k$    (the photons) of the $U(1)^{N-1}$   gauge group,   the  fermions listed  in Table~\ref{Simplest2} and the ``pion",    $\pi$.    The effective low-energy Lagrangian   has the form, 
\be    {\cal L}^{(eff)} =   {\cal L}(\psi, \eta,  A_{\mu}^{(i)} )  +   {\cal L}(\pi)    -    {\cal V}(\pi,  \psi, \eta)    + \ldots  \;,   \label{effective}   \ee
where $\psi, \eta$ are the fermions in Table~\ref{Simplest2}.     ${\cal L}(\psi, \eta,  A_{\mu})$ is the Lagrangian of the $U(1)^{N-1}$ theory with "electrons" $\psi, \eta$,
minimally coupled to the $U(1)^{N-1}$  gauge fields. 
 ${\cal L}(\pi) $  is the Lagrangian containing only the pion.
 Our task now is to learn about  $ {\cal L}^{(eff)} $ of  (\ref{effective}) as much as we can from symmetries, either broken, unbroken,   anomalous or non anomalous.
 
  In particular,  upon condensation the composite scalar field   $\psi\chi$  can be written as  
\be    U(x) =    (\psi\chi)^1_1(x)  =    \const. \,\Lambda^3   \, e^{i \pi(x)  / F}       \label{Ufield} 
\ee 
 where $F$ is the analogue of the pion decay constant.   $ {\cal L}(\pi)  $  contains the kinetic term 
 \be     {\cal L}(\pi)  =   \de^{\mu} U(x)^{\dagger}   \de_{\mu}  U(x) + \ldots, 
 \ee
 with possible higher order terms.

 Let us recapitulate the symmetries and their low-energy realizations
\beq  SU(N) \times  \frac{  SU(8)_{\rm f} \times {\tilde U}(1)  \times  U(1)_{\psi\chi} }{   {\mathbbm Z}_N    \times  {\mathbbm Z}_{8/ N^*}}    \longrightarrow    \frac{\prod_{\ell=1}^{N-1}  U(1)_{\ell}  \times SU(8)_{\rm f} \times {\tilde U}(1)    }{ \prod_{\ell=1}^{N-1} {\mathbbm Z}_\ell \times {\mathbbm Z}_N  \times {\mathbbm Z_2}  }\;.
\label{fullsymbreakingAgain}
\eeq
A possible local  interaction Lagrangian, consistent with the symmetries of the system, (\ref{fullsymbreakingAgain}),    is  
\be   {\cal V}(\pi,  \psi, \eta)(x)     \sim  U(x)^{N-2}  \, \underbrace {\psi \cdots \psi}_{4}  \, \underbrace {\eta \, \eta \cdots \eta}_{8}   + h.c.   =   \const. \,    \pi \pi \ldots \pi \; \, \underbrace {\psi \cdots \psi}_{4}\,  \underbrace {\eta\, \eta \cdots \eta}_{8}  + h.c.   \;,       \label{infra}  
\ee
with  any number of   pions,  four $\psi$'s and eight $\eta$'s.   It is invariant under the full UV symmetries     $SU(N)_{\rm c} \times SU(8)_{\rm f} \times {\tilde U}(1)  \times  U(1)_{\psi\chi} $, 
and {\it a fortiori}     with the unbroken symmetries $ \prod_{\ell=1}^{N-1}  U(1)_{\ell}  \times  SU(8)_{\rm f} \times {\tilde U}(1)$ as well as some discrete symmetries (see below),   surviving  in the infrared.  
Such a Lagrangian is not invariant under the anomalous  $U(1)_{an}$.  

The structure  of (\ref{infra})  may be understood from    
the original multi-fermion 't Hooft  effective instanton potential,  e.g.,   
\be    \epsilon_{a_1 a_2 \ldots  a_8}   \underbrace{   (\psi\chi)^i_j    (\psi\chi)^j_k \ldots     (\psi\chi)^p_i }_{N-2}   \underbrace{  (\psi \eta^{a_1}  \eta^{a_2})  \ldots  ( \psi  \eta^{a_7}   \eta^{a_8} )}_{4}\;,    \label{thooft}  
\ee
where a possible (certainly not unique)  way to contract the color $SU(N)$ and the flavor $SU(8)$  indices in an invariant way  is shown. The idea is to realize these  symmetry properties in terms of the infrared degrees of freedom.    In particular, by replacing the  condensate  $\psi\chi$ with a slowly varying fields  $U(x)$ of (\ref{Ufield}), 
one arrives at   
(\ref{infra}).

The Yukawa interactions among  $\pi,  \psi,\eta$ are forbidden by the  unbroken symmetries,  see Table~\ref{Simplest2}.     The only possible interactions among them 
are  those   arising   from the instanton-induced amplitude such as (\ref{infra}).    

The $\psi\chi\eta$  system has three global   $U(1)$ symmetries.    
${\tilde U}(1)$  symmetry is nonanomalous and remains unbroken.  It is a manifest symmetry of the low-energy effective action.  
The consequences of the nonanomalous but spontaneously broken $U(1)_{\psi\chi}$ symmetry and   anomalous but not-spontaneously-broken symmetry $U(1)_{an}$ 
are a little subtler.

$ U(1)_{\psi\chi} $  is spontaneously broken by the  $\psi\chi$  condensate.   It is a nonanomalous symmetry in the UV:    the fermion charges 
are such that  the   $ U(1)_{\psi\chi} $  anomalies due to the $SU(N)$  gauge interactions   
cancel. 
In the IR  it is spontaneously broken by the $\psi\chi$ condensate, a NG boson ($\pi$) is produced by the current from the vacuum,   and at the same time $SU(N)$  is dynamically broken to  $\prod_{\ell}  U(1)_{\ell}$.   

Now there seems to be a  paradox.   In the underlying (UV) theory   the  global  $U(1)_{\psi\chi}$ symmetry   acts on the fermions as:
\be    \psi\to  e^{i  \tfrac{N-2 }{N^* }    {\beta}} \psi\;, \qquad \chi \to   e^{- i \tfrac{N+2}{N^*}  {\beta}} \chi\;,    \qquad  \eta  \to \eta \;.         \label{and} 
\ee
The 't Hooft effective instanton potential  (\ref{thooft}) is indeed invariant under this. It is important to note however that such an invariance is not  invalidated by the $SU(N)$ anomalies as the contributions from the $\psi$ and $\chi$ fermions cancel,    see (\ref{current1}). 

Now in the infrared,     it is spontaneously broken and $U(1)_{\psi\chi}$ symmetry is realized partially nonlinearly,   as
\be        \pi(x)  \to    \pi(x)  -  \frac{4 F}{N^*}   \beta \;,\qquad U(x) \to  e^{-  \tfrac{4 i \beta}{N^*} }   \, U(x)\;,   \label{piontr}  
\ee
see   (\ref{Ufield} ) and (\ref{and}).    If we assume that the  fermions remaining massless in the infrared, see Table~\ref{Simplest2},   in particular    $\psi^{ii}$,  $i=1,2,\ldots, N$,
transform under   $U(1)_{\psi\chi}$   as in (\ref{and}):
\be          \psi^{ii}      \to  e^{i  \tfrac{N-2 }{N^* }    {\beta}} \psi^{ii}\;,   \label{psitr} 
\ee
the  effective potential in the infrared,   the first term of (\ref{infra}),   is  indeed  invariant.  
This shows that  $U(1)_{\psi\chi}$ symmetry  is realized partially nonlinearly ((\ref{piontr})) and partially linearly  ((\ref{psitr})) at low energies.  

But now  the  anomaly due to the   $\psi^{ii}$ loops,   
\be       \Delta   {\cal L}^{eff}  =     \frac{N-2}{N^*} \beta     \frac{1}{16 \pi^2}  \sum_{j=1}^{N-1}     e_{j}^2  \, F_{\mu \nu}^{(j)}    {\tilde F}^{(j)\,   \mu \nu}    \label{psianomaly}
\ee
cannot be cancelled, as there are no other massless fermions left in the infrared theory.  

Clearly such an argument is too na\"ive, and  neglects the fact that the  $U(1)_{\psi\chi}$   symmetry in the infrared does not only involve  the massless fermion, but also the pion,
transforming inhomogeneously as in 
(\ref{piontr}).  
The answer to this apparent puzzle is that   the low-energy  effective Lagrangian (\ref{effective})  actually contains an axion-like term 
\be           {\cal L}(\pi,  A_{\mu}^{(i)} )    =      \pi(x) \,       \frac{N-2}{4 F}  \frac{1}{16 \pi^2}  \sum_{j=1}^{N-1}     e_{j}^2  \, F_{\mu \nu}^{(j)}    {\tilde F}^{(j)\,   \mu \nu} 
\label{axionlike} 
\ee
which transforms under   (\ref{piontr})   as 
\be       \Delta   {\cal L}(\pi,  A_{\mu}^{(i)} )     =    -     \frac{N-2}{N^*} \beta     \frac{1}{16 \pi^2}  \sum_{j=1}^{N-1}     e_{j}^2  \, F_{\mu \nu}^{(j)}    {\tilde F}^{(j)\,   \mu \nu}  
\ee
cancelling exactly the anomaly due to the   $\psi^{ii}$ loops,  (\ref{psianomaly}), ensuring the    $U_{\psi\chi}(1)$ invariance of the system.

The conclusion is that the effective low-energy Lagrangian contains  an axion-like term, (\ref{axionlike}),  besides the  standard terms, explicit in  (\ref{infra}).  Another, equivalent way to reach the same conclusion  is to consider  various three-point functions, 
\be     \int \,d^4x \,  e^{i  q \cdot x}    \brc  0 | T \{   J_{\psi\chi}^{\mu}(x) A_{\ell}^{\nu}(y) A_{\ell}^{\lambda}(0)  \} | 0 \ckt\;,     \qquad \ell=1,2,\ldots, N-1\;,  
\ee
multiplying it by $q_{\mu}$ and  taking the limit  $ q_{\mu} \to 0$.     See Fig.~\ref{Cancel}

\begin{figure}
\begin{center}
\includegraphics[width=5in]{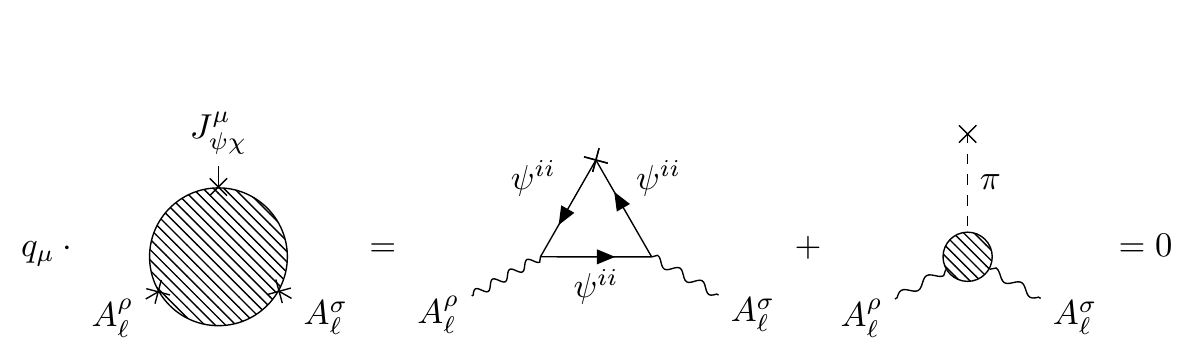}
\caption{\small   In UV, the absence of the gauge anomaly  in the  $U(1)_{\psi\chi}$ symmetry  
is due to the cancellation of  anomalies from  the $\psi$ and  $\chi$   fermions  (see (\ref{current1})).  
At low energies  it is secured by the cancellation between the 
contribution of the $\psi^{ii}$ loops  ($i=1,2,\ldots, N-1$)  and the pion pole term. 
%
  The     requirement   that the sum of the two cancel,   is nothing but the     $\pi \to  A_{\ell}  A_{\ell}$  amplitude   given by  (\ref{axionlike}).  
 }
\label{Cancel}
\end{center}
\end{figure}

Let us now consider  the  anomalous, but unbroken  $U(1)_{an}$  symmetry.     
As we noted already    ${\cal L}^{(eff)}$  is not invariant under  it.  The effect of the $U(1)_{an}$   anomaly however is  not  exhausted  in the explicit breaking  of  $U(1)_{an}$  symmetry in  $ {\cal V}(x)$. 
As   the  $U(1)_{an}$  charge of the low-energy, massless fermions is well defined,  see (\ref{current3}) and Table~\ref{Simplest2},    
 it  manifests itself   also  through  the 
    massless    $\psi$    fermion loops,  
    \be        J_{\rm an}^{\mu}  =      i \sum_{i=1}^N   {\bar \psi}_{ii}   {\bar \sigma}^{\mu}   \psi^{ii}  \;, \qquad    
\de_{\mu}   J_{\rm an}^{\mu}  =     \frac{1}{16 \pi^2}  \sum_{j=1}^{N-1}     e_{j}^2  \, F_{\mu \nu}^{(j)}    {\tilde F}^{(j)\,   \mu \nu}  \;.\label{U1anomalies}
\ee
Such an anomaly has a   natural interpretation   as   
a remnant  of  the original strong anomaly  (\ref{current3})   in the UV theory.     The original strong anomaly divergence equation has turned into the 
anomalous divergences due to the weak  $U(1)^{N-1}$ gauge interactions of the low-energy theory.

 To summarize, the symmetry realization pattern  of various $U(1)$ symmetries in the $\psi\chi\eta$ model is subtly  different  from  the one   in   QCD (the $U(1)_A$ problem and the massive $\eta$ meson) \cite{WittenU1}-\cite{Nath},  in QCD with the electromagnetic interactions  ($\pi_0 \to 2 \gamma$ decay through the ABJ anomaly),  or    in   
QCD with  Peccei-Quinn symmetry \cite{PQ,Kim,SVZ}  (with an extra scalar or heavy quarks,  giving rise to the axion and its coupling to the topological density of QCD), even though, here and there,  
we  see  some analogous  features.

\section{The generalized anomalies  \label{mixed}  }

The assumption of dynamical Abelianization  and its possible consequences studied in Sec.~\ref{fulla} 
are certainly consistent with the conventional 't Hooft anomaly matching requirement, as reviewed above.  
We now check such physics scenario in the $\psi\chi\eta$ model, against more stringent consistency requirements  arising from the 
mixed anomalies involving some higher symmetries.  
Let us recapitulate the symmetry of the system, 
\be    \frac{ SU(N) \times  {\tilde U}(1)  \times U(1)_{\psi\chi}  \times SU(8) }{   {\mathbbm Z}_{N}  \times   {\mathbbm Z}_{8/ N^*}  } \;, \label{fulluse}
\ee
where 
\bea        U(1)_{\psi\chi} &:& 
 \quad   \psi\to  e^{i  \tfrac{N-2 }{N^* }    {\beta}} \psi\;, \quad \chi \to   e^{- i \tfrac{N+2}{N^*}  {\beta}} \chi\;, \nonumber \\  
         {\tilde U}(1)&:&  \quad    \psi \to  e^{ 2 i \gamma} \, \psi\;, \quad  \chi \to  e^{-  2 i \gamma}\, \chi\;, \quad  \eta  \to  e^{- i \gamma} \, \eta   \;,
\label{Inacc} \eea
and 
\bea    {\mathbbm Z}_{8/N^*}  &=&    SU(8)    \cap  ( {\tilde U}(1)  \times     U(1)_{\psi\chi} )\;, \nn \\   {\mathbbm Z}_{N}   &=&  SU(N)    \cap  {\tilde U}(1) \;.\label{redundancy}
\eea
We wish now to find out if   new, stronger consistency conditions on the realization of these  symmetries 
arise by making full use of the global structure of the symmetry group,  Eq.~(\ref{fulluse}), i.e.,  by gauging some  
1-form center symmetries,  such as  ${\mathbbm Z}_{N}$ and/or  ${\mathbbm Z}_{8/N^*}$. 
 To do so, however, it is necessary to make use of  symmetries which are not broken by the color $SU(N)$ or  the weak $SU(8)$ gauge interactions,
including the nonperturbative effects (instantons).  
Gauging the $1$-form   ${\mathbbm Z}_{8/N^*}$ symmetry involves necessarily gauging the ${\tilde U}(1)$ symmetry, which is
already broken by the $SU(8)$ instantons, see Eq.~(\ref{Inacc}).    It would not be a simple task to disentangle the effects of the new 
anomalies due to the gauging of a  discrete  center symmetry,  from the  conventional anomalies due to the $SU(8)$ instantons.  Such a precaution appears to be relevant,
because we are here interested in possible new mixed anomalies on continuous symmetries    ${\tilde U}(1)$ and  $U(1)_{\psi\chi}$ or on some of their discrete subgroups. 

These considerations lead us to preclude the idea of gauging the 1-form   ${\mathbbm Z}_{8/N^*}$ symmetry, and  below we shall focus on the color-flavor locked 1-form  ${\mathbbm Z}_{N}$ center symmetry,   gauge it in conjunction with   some 0-form $U(1)$ symmetries of the model, and  examine whether such a simultaneous  gauging would suffer from some topological obstructions  (generalized   't Hooft anomalies).

\subsection{Calculation of the  mixed anomalies}

The global structure of the symmetry of the gauge groups reproduced in  (\ref{fulluse}) $\sim$  (\ref{redundancy}) means that 
it should be possible to introduce a more faithful way the redundancies in the summation over the gauge field configurations are eliminated.  
In our context, this brings us to consider effectively a projective  $SU(N)/{\mathbbm Z}_{N}$ group as the strong gauge group, and with
its  2-form   $B_{\rm c}^{(2)}$  fields carrying  a fractional   't Hooft fluxes
      \be      \frac{1}{8\pi^2}  \int_{\Sigma_4}    (B^{(2)}_\rmc)^2  \in   \frac {{\mathbbm Z}}{N^2}\;.  \label{fraction} 
      \ee
The way this is done concretely has been explained in \cite{GKSW}\sp\cite{Anberhongson}, by introducing the  1-form   ${\mathbbm Z}_N$
gauge field, and imposing the condition (\ref{condition})  for a  ${\mathbbm Z}_N$ gauge field.  As our  ${\mathbbm Z}_{N}$ center symmetry
is a color-flavor locked symmetry, to render  it properly a  1-form symmetry one must accompany the $SU(N)$ Wilson loop with a ${\tilde U}(1)$ holonomy  (Aharonov-Bohm) loop for the fermions.   See also the related discussion (vii) in Sec.~\ref{sec:observations}  below.

     Taking these points into account   now we  introduce the  gauge fields   as 
\begin{itemize}
  \item     $a_{\rm c}$:  ~~   the $SU(N)$ color gauge field;
  \item   $a_{\rm f}$:  ~~   the $SU(8)$   flavor gauge field;
  \item    ${\tilde A}$:  ~~   the gauge field for  ${\tilde U}(1)$;
    \item    ${A}_{\psi\chi}$:  ~~   the gauge field for  $U(1)_{\psi\chi}$;
    \item   $B_{\rm c}^{(1)}$,  $B_{\rm c}^{(2)}$: ~~       ${\mathbbm Z}_N$ gauge field.
\end{itemize}
The  pairs of  gauge fields    $\big(B_{\rm c}^{(1)}, B_{\rm c}^{(2)}\big) $    for the  ${\mathbbm Z}_N$     1-form symmetry    satisfy the constraints\footnote{The suffices are to indicate the 1-form or 2-form nature of these gauge fields,  e.g., $ B^{(1)} =  B_{\mu} dx^{\mu}$, 
 etc.   }
\be      N  \,B_{\rm c}^{(2)} =   d B_{\rm c}^{(1)}\;.   \label{condition}
\ee 
Following the by now well-understood procedure for gauging  a 1-form discrete symmetry,  one also introduces redundant 
$U(N)$  gauge fields 
\begin{itemize}
  \item     ${\tilde a}_{\rm c}$:  ~~   the $U(N)$ color gauge field;
\end{itemize}  
 where
 \be
\tilde{a}_{\rm c}=a_{\rm c}+\frac{1}{N}B^{(1)}_{\rm c}\;.
\ee
The central idea is that one then 
 imposes  the invariance under the $1$-form gauge transformations 
\bea    & B_{\rm c}^{(2)}   \to  B_{\rm c}^{(2)} + d \lambda_{\rm c}\;, \qquad    B_{\rm c}^{(1)}   \to  B_{\rm c}^{(1)} + N   \lambda_{\rm c}\;,    & \label{gauge1} \nn \\ 
  & {\tilde a}_{\rm c} \to   {\tilde a}_{\rm c} +   \lambda_{\rm c} \;, & \label{gauge3}
\eea
where    $\lambda_{\rm c}$ is a   1-form  $U(1)$  gauge function.\footnote{In the standard gauging of a $0$-form $U(1)$ symmetry,
$\psi \to e^{i \lambda}\psi$;  $A_{\mu}  \to A_{\mu} +  \frac{i}{e} \partial_{\mu} \lambda$,  $\lambda(x)$ is a $0$-form gauge function. } 
The ${\tilde U}(1)$ and $U(1)_{\psi\chi}$ gauge fields    ${\tilde A}$ and $A_{\psi\chi} $   transform under these  as  (see (\ref{ZN}),(\ref{def:U1psichi}))
\be           {\tilde A}  \to {\tilde A} - \lambda_{\rm c}\;,\qquad     A_{\psi\chi}  \to      A_{\psi\chi}\;.  \label{gauge4}
\ee
The requirement of the invariance under the $1$-form gauge transformations (\ref{gauge1})-(\ref{gauge4})  realizes the elimination of the redundancies,  (\ref{redundancy}).

The (1-form) gauge invariant Dirac operators  are  accordingly
\be   d +    \calR_{\rmS} ({\tilde a}_{\rm c} -  \frac{1}{N} B_{\rm c}^{(1)})  +     2\, ({\tilde A} + \frac{1}{N}  B_{\rm c}^{(1)} )   +   \frac{N-2}{N^*}   A_{\psi\chi}   \;,   \label{kinet1}
\ee
acting on $\psi$,  
\be   d  +   \calR_{\rm A^*}({\tilde a}_{\rm c} -  \frac{1}{N} B_{\rm c}^{(1)})    -  2   ({\tilde A} + \frac{1}{N}  B_{\rm c}^{(1)} )  -    \frac{N+2}{N^*}  A_{\psi\chi}   \;,    \label{kinet2}
\ee
acting on $\chi$,   and  
\be   d - ({\tilde a}_{\rm c}  -    \frac{1}{N} B_{\rm c}^{(1)})    +  a_{\rm f} 
  -    ({\tilde A} + \frac{1}{N}  B_{\rm c}^{(1)})    \;, \label{kinet3}
\ee
acting on $\eta$.   Note that written this way the expression inside each bracket is invariant under  (\ref{gauge1})-(\ref{gauge4}).
In the above we have introduced a (hopefully) self-evident notation for $SU(N)$ algebras in symmetric  and anti-antisymmetric representations adequate for the $\psi$ and $\chi$ fields.  By construction  the combination ${\tilde a}_{\rm c} -  \frac{1}{N} B_{\rm c}^{(1)}$  belongs to  the $SU(N)$ algebra.

Before proceeding, it is useful to record the relation  between   ${\tilde A}$, ${A}_{\psi\chi}$  and    the straightforward   $A_{\psi}$, $A_{\chi}$, $A_{\eta}$ gauge fields associated with the 
$U(1)_{\psi}$, $U(1)_{\chi}$, $U(1)_{\eta}$ fermion number symmetries:
\be    \psi\to e^{i \alpha_{\psi}}\psi\;, \qquad   \chi \to e^{i \alpha_{\chi}}\chi\;, \qquad   \eta \to e^{i \alpha_{\eta}} \eta\;. 
\ee
They can be read off from (\ref{kinet1})-(\ref{kinet3}):
\be    A_{\psi} =     2\, {\tilde A}    +  \frac{N-2}{N^*}  A_{\psi\chi}  \;,\qquad    A_{\chi} =     -  2\, {\tilde A}    -    \frac{N+2}{N^*} A_{\psi\chi}  \;, \qquad     A_{\eta} =   -    {\tilde A}\;.   \label{translate} 
\ee

The  gauge field tensors felt by the fermions corresponding to   (\ref{kinet1})-(\ref{kinet3})    are:
\bea &&  \calR_{\rmS}(F(\widetilde{a}) - B_{\rm c}^{(2)})    +      2\, (d {\tilde A} +   B_{\rm c}^{(2)} )    +   \frac{N-2}{N^*}      d  A_{\psi\chi}     \;,  \nonumber \\
&& \calR_{\rm A^*}(F(\widetilde{a}) - B_{\rm c}^{(2)})      -   2\, (d {\tilde A} +   B_{\rm c}^{(2)}  )   -   \frac{N+2}{N^*}     d  A_{\psi\chi}      \;, \nonumber \\
&&      \calR_{\rm F^* }(F(\widetilde{a}) - B_{\rm c}^{(2)})    +     F_{\rm f}(a_{\rm f})        -   \, (d {\tilde A} +   B_{\rm c}^{(2)} )   \;. \phantom{\frac{1}{2}}
\eea
 The anomalies are compactly  expressed by a $6$D anomaly functional
\bqa {\cal A}^{6D} &=&   \int \frac{2\pi}{3!  (2\pi)^3}     \Big\{   {\tr}_{\rm c}   \left(   \calR_{\rmS} ({\tilde F}_{\rm c} -  B_{\rm c}^{(2)} ) + 2\, (d {\tilde A} +   B_{\rm c}^{(2)} )  +  \frac{N-2}{N^*}   d  A_{\psi\chi}     \right)^3  \nonumber \\
  & &   \qquad \qquad  + \, {\tr}_{\rm c}   \left( \calR_{\rm A^*}({\tilde F}_{\rm c} -  B_{\rm c}^{(2)} )  -   2\, (d {\tilde A} +   B_{\rm c}^{(2)} ) -  \frac{N+2}{N^*}   d  A_{\psi\chi}     \right)^3  \nonumber \\
  & & \qquad   \qquad + \,    {\tr}_{c,8}   \left(  - ({\tilde F}_{\rm c} -  B_{\rm c}^{(2)} ) +     F_{\rm f}(a_{\rm f})     -   \, (d {\tilde A} +   B_{\rm c}^{(2)})    \right)^3   \Big\} \;.       \label{anomaly6D}
\eea

Expanding the 6D anomaly functional  (\ref{anomaly6D}),  one finds
 \bqa  & &    \frac{2\pi}{3!  (2\pi)^3}  \int  \big\{  [(N+4) - (N-4) -8 ]   \,  {\tr}_{\rm c}  ({\tilde F}_{\rm c} -  B_{\rm c}^{(2)} )^3  \big\}   \nonumber      \\ 
  & &   + \,     \frac{2\pi   N }{3!  (2\pi)^3}   \int    {\tr}_{8}    \,     (F_{\rm f}(a_{\rm f}))^3  \nonumber \\ 
 & &   + \,  \frac{1}{8 \pi^2}   \int    {\tr}_{\rm c}  ({\tilde F}_{\rm c} -  B_{\rm c}^{(2)} )^2     \big\{   (N+2)  [ 2\, (d {\tilde A} +   B_{\rm c}^{(2)})  +   \frac{N-2}{N^*}  d  A_{\psi\chi}]   \nonumber \\
 & &   \qquad \qquad + \,   (N-2) [ -  2\, (d {\tilde A} +   B_{\rm c}^{(2)} )   -  \frac{N+2}{N^*}    d  A_{\psi\chi} ]  + 8\,   [   -   \, (d {\tilde A} +   B_{\rm c}^{(2)} )  ]   \big\}     \nonumber \\
 & &   + \,   N\,   \frac{1}{8 \pi^2}   \int     {\tr}_8    \, (F_{\rm f}(a_{\rm f}))^2    [   -   \, (d {\tilde A} +   B_{\rm c}^{(2)})  ]  \nonumber \\
 & &   + \, \frac{1}{24 \pi^2}   \int   \big\{  \frac{N(N+1)}{2}  [  2\, (d {\tilde A} +   B_{\rm c}^{(2)})  +   \frac{N-2}{N^*}  d  A_{\psi\chi}]^3 \nonumber \\ 
 & &    \qquad \qquad \ + \,   \frac{N(N-1)}{2}  [-  2\, (d {\tilde A} +   B_{\rm c}^{(2)} )   -  \frac{N+2}{N^*}  d  A_{\psi\chi} ]^3    \nonumber \\
 & &    \qquad \qquad \ + \,  8 N  [    -   \, (d {\tilde A} +   B_{\rm c}^{(2)} )   ]^3       \big\}  \;,\label{third}
 \eea
by  making use of the known formulas for the traces of  quadratic and cubic forms in different representations.
The terms in the first line, proportional to   $ {\tr}_{\rm c}  ({\tilde F}_{\rm c} -  B_{\rm c}^{(2)} )^3 $   trivially cancel out, reflecting the anomaly-free nature of the  $SU(N)$ color group. 
 Note also that the third and fourth lines, namely  the terms containing   $  {\tr}_{\rm c}  ({\tilde F}_{\rm c} -  B_{\rm c}^{(2)} )^2$,    
completely cancel each other, due to the fact that only anomaly-free  combinations (${\tilde U}(1)$ and $U(1)_{\psi\chi}$) of the $U(1)$ symmetries are being considered.   A further gauging of the 1-form center symmetry (by the introduction of  $B_{\rm c}^{(2)}$) obviously does not affect this.  
 Taking into account these cancellations one arrives at 
  \bea   &&   \frac{2\pi   N }{3!  (2\pi)^3}   \int    {\tr}_{8}    \,     (F_{\rm f}(a_{\rm f}))^3 +
  N\,   \frac{1}{8 \pi^2}   \int     {\tr}_8    \, (F_{\rm f}(a_{\rm f}))^2    [   -   \, (d {\tilde A} +   B_{\rm c}^{(2)})  ]   \nonumber \\
 & &   + \,  \frac{1}{24 \pi^2}   \int  \big\{  \frac{N(N+1)}{2}  [  2\, (d {\tilde A} +   B_{\rm c}^{(2)} )  +   \frac{N-2}{N^*}  d  A_{\psi\chi}]^3 \nonumber \\ 
& &    \qquad \qquad \  \ + \frac{N(N-1)}{2}  [-  2\, (d {\tilde A} +   B_{\rm c}^{(2)})   -  \frac{N+2}{N^*}     d  A_{\psi\chi}  ]^3    \nonumber \\
& &    \qquad \qquad  \ \ + 8 N  [    -   \, (d {\tilde A} +   B_{\rm c}^{(2)} )   ]^3       \big\} \;. \label{also}
 \eea

\subsection{Observations \label{sec:observations} }

The mixed  anomalies involving  the $0$-form  ${\tilde U}(1)$ and  $U(1)_{\psi\chi}$  symmetries and the $1$-from  discrete center symmetry ${\mathbbm Z}_N$ 
can now be found by studying  the terms
\be      \propto    B_{\rm c}^{(2)}  \,  {\tilde A}\;,    \qquad     \propto    B_{\rm c}^{(2)}  \,  A_{\psi\chi} \;,  
\ee
in the $5$D  WZW  action, and considering the variations 
\be     \delta  {\tilde A}   =    d  \, \delta  {\tilde A}^{(0)}\;,   \qquad     \delta  A_{\psi\chi}    =    d  \, \delta  A_{\psi\chi}^{(0)}\;,
\ee
which correspond to the phase transformations of the fermions  (\ref{Inacc}),   with   
\be     \delta {\tilde A}^{(0)} =\gamma\;, \qquad  \delta A_{\psi\chi}^{(0)}=\beta\;, 
\ee
to give the anomalous variations of the partition function in the boundary  $4$D action (the anomaly inflow).
It turns out that  the anomaly expression (\ref{also})   contains quite a remarkable set of interesting physics implications.

\begin{description}

\item[(i)]  The first  line of (\ref{also})  simply  represents the  $SU(8)^3$ and ${\tilde U}(1) - [SU(8)]^2    $  anomalies,  dressed by the 2-form gauge field $B_{\rm c}^{(2)}$.
The associated matching of the conventional anomalies in the UV and IR has already been discussed in Sec.~\ref{matching}.  As noted  in \cite{BKL1},  once the standard  't Hooft anomaly matching equations are satisfied 
for {\it  continuous}   symmetries, the 1-form gauging (introducton of the $B_{\rm c}^{(2)} $  fields and their fractional fluxes) does not affect the UV-IR anomaly matching.  

  \item[(ii)] 
     The terms proportional to  $ (d {\tilde A} +   B_{\rm c}^{(2)} )^3$  in (\ref{also})  cancel each other completely. This means that the  mixed anomaly of the form 
   \be   d{\tilde A}   (B_{\rm c}^{(2)})^2  \ee
   is  absent.    There is no  obstruction in gauging the  ${\tilde U}(1)$ symmetry  together with the color-flavor locked ${\mathbbm Z}_N$  center symmetry. 
       The ${\tilde U}(1)$ symmetry may well remain unbroken.

  \item[(iii)]  The mixed anomaly of the form    $dA_{\psi\chi} (B_{\rm c}^{(2)})^2$ is present: it is given by 
  \be   -  \frac{4 N^2}{N^*}   dA_{\psi\chi} (B_{\rm c}^{(2)})^2\;,     \label{however}
  \ee
  which is equal to 
  \be      N^2    (dA_{\psi}+ d A_{\chi}) (B_{\rm c}^{(2)})^2\;,    \label{firstline} 
  \ee
  in view of Eq.~(\ref{translate}).  In other words,  $U(1)_{\psi\chi}$  symmetry cannot be gauged consistently, when  
  the 1-form color ${\mathbbm Z}_N$  symmetry is gauged.

 To the best of our knowledge,   the last phenomenon in  is new. 
   In all studies on generalized 't Hooft anomaly matching studied  so far \cite{GKSW}\sp\cite{Wan:2018djl},   nontrivial mixed anomalies concerned the possible breaking of some discrete symmetry.   
Here,  we find that a {\it continuous}  $U(1)$ symmetry is affected  by a mixed $0$-form-$1$-form anomaly.   It appears that this is a typical, rather than exceptional, phenomenon in chiral gauge theories.

  \item[(iv)]  The mixed  anomaly  (\ref{firstline})  shows also that $ ({\mathbbm Z}_{N+2})_{\psi}$, $({\mathbbm Z}_{N-2})_{\chi}$, 
  are both broken by the 1-form gauging  of    ${\mathbbm Z}_N$.   The action of a $({\mathbbm Z}_{N+2})_{\psi}$ transformation,  for instance,  is described by the variation in the $5$D action 
  \be  \delta A_{\psi} =  d  \, \delta A_{\psi}^{(0)}\;, \qquad      \delta A_{\psi}^{(0)} =  \frac{2\pi  k } {N+2}\;; \qquad   k \in {\mathbbm Z}\;
  \ee
  which then yields the  anomalous variation of  the $4$D action  
   \be   \delta S =    \frac{1}{8\pi^2}  \int_{\Sigma_4}   N^2    (B^{(2)}_\rmc)^2      \frac{2\pi  k } {N+2}  =      \frac{2\pi  k } {N+2}     {\mathbbm Z}\;,
      \ee
      under \be  \psi \to e^{\frac{2\pi i k}{N+2}} \psi\;,  \ee  
      where the fractional 't Hooft  flux   (\ref{fraction})  
     of our  $SU(N)/{\mathbbm Z}_{N}$ theory has been taken into account.
     
  Note that even though a generic $({\mathbbm Z}_{N+2})_{\psi}$ transformation changes the partition function,   the effect of 
$k=N+2$  transformation is found to be trivial.  This is as it should be. By definition  a   $({\mathbbm Z}_{N+2})_{\psi}$ ``transformation" with  $k=N+2$  means $\psi \to \psi$:  it is not a transformation at all. 
    In other words,  the coefficient $N^2$  found above, (\ref{firstline}), is significant.

  Similarly for  $({\mathbbm Z}_{N-2})_{\chi}$.

\item[(v)]     However,  a particular  subgroup, 
 \be     {\mathbbm Z}_{N^*}\subset   ({\mathbbm Z}_{N+2})_{\psi}\times   ({\mathbbm Z}_{N-2})_{\chi}\;,   \label{definedcosi1}
  \ee
 remains unaffected by the mixed anomaly  involving the  color-flavor locked   1-form
 ${\mathbbm Z}_N$ symmetry.   
   This can be seen as follows.   The condition
  imposed for the conservation by   (\ref{firstline})    is that 
  \be \frac{ 2\pi k}{N+2} +  \frac{ 2\pi \ell }{N- 2} = 2\pi \times {\rm integer}\;.     \label{discrete}
  \ee
  Such a condition can be solved by 
  \be     k=  \frac{N+2}{N^*} p\;, \quad  \ell=  \frac{N-2}{N^*}  (N^*-p)\;,  \qquad    p =1,2,.\ldots, N^*\;,
  \ee
  that is
  \be   \psi  \to  e^{  2 \pi i  \, p / N^*} \psi\;; \quad \chi \to    e^{ -    2  \pi  i   \,  p / N^*} \chi\;, \qquad p \in {\mathbbm Z}_{N^*}\;, \quad N^* = GCD(N+2, N-2)  \;.         \label{this} 
  \ee

Note that the  conservation of  this  ${\mathbbm Z}_{N^*}$  subgroup is  perfectly consistent with the  assumption that condensation 
  $\brc \psi \chi \ckt \ne 0 $ forms in the infrared, as was noted  in  Sec.~\ref{LE}, see  (\ref{znspsichi}).

\item[(vi)]       We   saw above (the point (iii)) that  
  $U(1)_{\psi\chi}$  symmetry cannot be gauged consistently, when  
  the 1-form color ${\mathbbm Z}_N$  symmetry is gauged.    However the form of the mixed anomaly    (\ref{however})
shows   that a global $U(1)_{\psi\chi}$ transformation gives a trivial phase to the partition function,    for its discrete subgroup, $\mathbbm Z_{4/N^*}\subset U(1)_{\psi\chi}$, as defined in (\ref{Z4N1}), (\ref{Z4N2}). 
  This is perfectly consistent with the dynamical Abelianization, as $<\psi\chi>\neq 0$ implies $U(1)_{\psi\chi}\rightarrow \mathbbm Z_{4/N^*}$, thus no residual anomaly matching condition has to be satisfied \footnote{We recall  that $\mathbbm Z_{4/N^*}\subset U(1)_{\psi\chi}$    
  is also a subgroup of $SU(8)\times \tilde U(1)$, thus it is naturally included in the global IR group as written in (\ref{fullsymbreaking}).
}.

 \item[(vii)]  
To understand better the situation, it is useful to see how this anomaly arises from the fractionalization of the $\tilde U(1)$ fluxes and the more mundane $U(1)_{\psi\chi}-[\tilde U(1)]^2$ anomaly. 
As one gauges $\mathbbm Z_N^{(1)}$  (i.e., ``1-form $\mathbbm Z_N$ symmetry"), i.e. one  considers $\frac{SU(N)_c\times \tilde U(1)}{\mathbbm Z_N}=U(N)$ gauge bundles that are not $SU(N)_c\times \tilde U(1)$ gauge bundles, both the instanton number and the fluxes of $\tilde U(1)$ are fractionalized.
The $U(1)_{\psi\chi}-[SU(N)_c]^2$ (strong) anomaly vanishes identically, therefore the fractionalization of the $SU(N)_c$ instanton number has no consequences on  $U(1)_{\psi\chi}$. 
However, the $U(1)_{\psi\chi}-[\tilde U(1)]^2$ anomaly does not vanish. In particular, by gauging $\tilde U(1)$ but not $\mathbbm Z_N^{(1)}$, one sees from the $U(1)_{\psi\chi}-[\tilde U(1)]^2$ anomaly that the partition function gets a phase,
\be \mathcal{Z} \rightarrow e^{-i\beta \frac{4N^2}{N^*}\int d\tilde A \wedge d \tilde A}\mathcal{Z}\;,\ee
under $e^{i\beta}\in U(1)_{\psi\chi}$. This phase is trivial ($2\pi \mathbbm Z$) for $\mathbbm Z_{4N^2/N^*}\subset U(1)_{\psi\chi}$. 
By gauging also $\mathbbm Z^{(1)}_N$, as the $\tilde U(1)$ fluxes fractionalizes, 
\be \frac{1}{2\pi}\int d\tilde A = \mathbbm Z \rightarrow \frac{1}{2\pi}\int d\tilde A + B^{(2)}_c = \frac{1}{N}\mathbbm Z\;, \ee
the 't Hooft anomaly free subgroup of $U(1)_{\psi\chi}$ is further reduced to $\mathbbm Z_{4/N^*}$.

	\item[(viii)] The fact that $\mathbbm Z_{4/N^*}$ is free of mixed anomalies is, by itself, an interesting consistency check. This is  because, being $\mathbbm Z_{4/N^*} \subseteq SU(8)\times \tilde U(1)$, as $SU(8)\times \tilde U(1)$ does not suffer any mixed 't Hooft anomaly with $\mathbbm Z^{(1)}_{N}$, also $\mathbbm Z_{4/N^*}$  must be free of such anomaly. Instead, from the calculation above, the fact that $\mathbbm Z_{4/N^*}$ is free of mixed anomaly is nontrivial: if the coefficient of the $dA_{\psi\chi}\left(B_c^{(2)}\right)^2$ term in (\ref{also}) were different from $-\frac{4N^2}{N^*}$,
 it would not hold.

  \item[(ix)]  $({\mathbbm Z}_{8})_{\eta}$  and $SU(8)$ itself, are neither broken by the standard instantons nor in the presence of the 
  1-form gauge fields $\big(B_{\rm c}^{(2)}, B_{\rm c}^{(1)}\big)$.    
    
\end{description}

\section{Summary and Discussion}

In this work,  we revisited the infrared dynamics of the chiral $\psi\chi\eta$ theory,  assuming dynamical Abelianization 
caused by bifermion condensate in the adjoint representation of the $SU(N)$ gauge group. In the first part, the symmetries of the system are 
studied and the working of the conventional 't Hooft anomaly matching has been briefly reviewed, and the possible form of the effective low-energy action 
is studied,  by taking also into account  also of the strong anomaly.  

In the second part of the work, we have checked these ideas against more stringent 
constraints following the mixed-anomaly involving certain  0-form  $U(1)$ symmetries and 1-form color-flavor locked ${\mathbbm Z}_N$ center symmetry. The results of the analysis,
summarized in Sec.~\ref{sec:observations},   tell us that the proposed infrared physics, characterized by dynamical Abelianization, is  consistent with the implications of the  mixed anomalies and, perhaps, implied by them. 
The comparison between the  implications of the mixed anomalies and those expected from the assumption of  the bifermion adjoint condensate  and  dynamical Abelianization,   is shown  in Table~\ref{symmetries}. 
   It is seen that the 
pattern of the symmetry realization (breaking)  in the infrared, suggested by the   mixed anomalies involving the gauged  1-form ${\mathbbm Z}_N$ symmetry, are well reproduced by the 
dynamical Abelianization proposed in this work.  
\begin{table}[h]
  \centering 
  \begin{tabular}{|c |c|c| c|c|c|c|c|c| }
\hline
$ \phantom{{{   {  {\yng(1)}}}}} \!  \! \! \! \! \!\!\!$ &                $  {\tilde U}(1)   $   &     $  U(1)_{\psi\chi}   $   &    $({\mathbbm Z}_{N+2})_{\psi}$ &  $({\mathbbm Z}_{N-2})_{\chi}$       &    $SU(8)_{\eta}$      &   ${\mathbbm Z}_{N^*}$     &     $  {\mathbbm Z_{4/N^*}}$
     \\   
 \hline 
 Mixed Anomalies &     \checkmark   &  X  &   X  &   X  &         \checkmark    &      \checkmark   &      \checkmark     \\
 \hline
   Dyn. Abel.     &     \checkmark   &  X &    X &   X  &      \checkmark    &    \checkmark    &     \checkmark    \\
 \hline
\end{tabular}  
\caption{\small    Dynamical Abelianization postulate of the present work is confronted with the implications of the mixed anomalies.
      $\checkmark $   for a conserved symmetry,  X     for a broken symmetry.  The discrete  ${\mathbbm Z}_{N^*}$   symmetry is defined in     (\ref{znspsichi}), or  in 
      (\ref{definedcosi1})-(\ref{this}).       $  {\mathbbm Z_{4/N^*}}$  is defined in     (\ref{Z4N1}).
    }\label{symmetries}
\end{table}

In this work  we have examined the consistency of the hypothesis of dynamical Abelianization, that a bifermion condensate forms in the infrared, of the form,  (\ref{psichicond}),  (\ref{psichicondBis}). It is possible that a bifermion condensate in the adjpoint representation forms,  but with a different symmetry breaking pattern, e.g.,  
\be
		 \brc (\psi\chi)^i_j \ckt = c\;\begin{cases} (N-m)\; \delta^i_j \;, \qquad i, j=1, ..., m\\
			-m\;  \delta^i_j  \;,   \qquad     i, j=m+1, ..., N
		\end{cases} \quad c \sim \mathcal{O}(\Lambda_0^3)\;.
		\ee
In this case,  the strong gauge group would be broken as  
		\be
		SU(N)_{\rm c}  \rightarrow SU(m)_{\rm c} \times SU(N-m)_{\rm c'}\times U(1)_e\;,
		\ee
		where $U(1)_e$ is generated by $T \propto {\rm diag} (\underbrace{N-m}_{n}, -\underbrace{m}_{N-m})$.
		A quick look at the massless  spectrum expected from such a  symmetry breaking shows that the system below the scale $\Lambda_0$, is basically a pair of  $\psi\chi\eta$  models with  $SU(m)$ and  $SU(N-m)$ gauge groups, respectively. The system is asymptotically free and continues to evolve towards the infrared.  We shall not pursue further such   
a tumbling-like scenario, but it is possible that at the end the system flows into the full dynamical Abelianization, studied in Sec.~\ref{fulla}.

Even though  we have focused our attention  in this work on the $\psi\chi\eta$  theory for definiteness,   there are other chiral gauge theories  in which  a  bifermion condensate in the adjoint representation might occur  and  in which  dynamical Abelianization 
 might be decisive in determining  the infrared physics. Possible examples are 
\begin{description}
  \item[(i)]    $SU(N)$  theory  (with $N$ even), with odd number of fermions in the self-adjoint antisymmetric  order $N/2$  tensor representation, 
  studied in \cite{Yamaguchi,BKL1}\;;  
   
  \item[(ii)]   A generalization of the    $SU(N)$    $\psi\chi\eta$  model with a matter  fermion content,
  \beq
   \psi^{\{ij\},\, m}\;, \qquad \chi_{[ij]} \;, \qquad  \eta^B_j\;,  \qquad    m=1,2,\,\quad  B=1,2,\ldots, N + 12\;,
\eeq
or
\be      2 \,\,  \yng(2) +  {\bar {\yng(1,1)}} +   (N+12) \, {\bar {\yng(1)}}\;. 
\ee
studied in \cite{BK}\;,   
and 
  \item[(iii)]   $SU(N)$  theories with  fermions  in the complex representation,  $ \tfrac{N-4}{k}  $   $\psi^{\{ij\}}$'s and    $\tfrac{N+4}{k}$   ${\bar  \chi_{[ij]} }$,
   \be   \frac{N-4}{k}  \,\,   \yng(2)   \oplus   \frac{N+4}{k}     \,\,    {\bar  {\yng(1,1)} }  \ ,   \ee
($k$ being a common devisor of $N-4$ and $N+4$)   studied recently    \cite{BKL1,Anberhongson}.  
  \end{description}
In all of them, the conventional 't Hooft anomaly matching analysis is consistent with  dynamical Abelianization hypothesis,    
and in some cases, the preliminary analysis  involving the generalized symmetries and the mixed anomalies appears to  give  further  support   \cite{BKL1,Anberhongson} for it.
Still, in some of this class of models,  the symmetry breaking pattern may be different from dynamical Abelianization, allowing for  a more general  types of  infrared gauge theories.   We will come back to the discussion of these models in a separate investigation.

\section*{Acknowledgment} 

This work is supported by the INFN special initiative grant, GAST (``Gauge and String Theories").

\appendix

\section{Chiral Ward-Takahashi identities and  NG bosons \label{WT}}

Let us briefly review the fate of a continuous, global symmetry, say   $G_{\rm f}$, when a condensate forms in the infrared  which is not invariant under it. 
Let  the associated conserved current be  $J_{\mu}$ and charge $Q$.   The field $\phi$ (elementary or composite) condenses and breaks  $G_{\rm f}$.  
The field  ${\tilde \phi}$  (elementary or composite)   is such that it is transformed by the  $G_{\rm f}$  transformation  into   $\phi$:
\be  Q \equiv  \int d^3x    J_{0}\;, \qquad    [Q,  {\tilde \phi}] =    \phi\;, \qquad \brc \phi \ckt  \ne 0\;.   \label{WTI1}
\ee
The Ward-Takahashi like identity  
\bea  && \lim_{q_\mu \to 0 }  i q^{\mu} \int d^4x   \, e^{- i q \cdot x} \brc  0| T\{ J_{\mu}(x) \,  {\tilde \phi}(0) \}  | 0 \ckt =    \lim_{q_\mu \to 0 }  \int d^4x   \, e^{- i q \cdot x}    \de_{\mu}   \brc  0| T\{ J_{\mu}(x) \,  {\tilde \phi}(0) \}  | 0 \ckt =  \nonumber \\
&&=   \int d^3x    \brc  0| [  J_{0}(x),   {\tilde \phi}(0)  ]  | 0 \ckt =    \brc  0| [  Q,   {\tilde \phi}(0) ]  | 0 \ckt =   \brc  0| \phi(0) | 0 \ckt  \ne 0 \;.  \label{chWT}
\eea
implies that the two-point function
\be    \int d^4x   \, e^{- i q \cdot x} \brc  0| T\{ J_{\mu}(x) \,  {\tilde \phi}(0) \}  | 0 \ckt  \label{2point} 
\ee
is singular at $q^{\mu}  \to 0$.  Under the assumption that  the $G_{\rm f}$ symmetry is broken spontaneously,   such a singularity is due to a massless scalar particle  
in the spectrum.   This particle, known as  Nambu-Goldstone (NG) boson   (a ``pion", symbolically)  must be produced from the vacuum by the broken current  $J^{\mu}$:
\be   \brc 0| J_{\mu}(q)  |  \pi \ckt =  i q_{\mu}  F_{\pi}\;, \qquad   \brc \pi | {\tilde \phi} | 0 \ckt \ne 0\;.     \label{WTI2}
\ee
such that 
the two point function  (\ref{2point}), when contracted by   $q^{\mu}$,    behaves as 
\be   \lim_{q^{\mu} \to 0} \,  q^{\mu} \cdot q_{\mu} \,  \frac{F_{\pi}   \brc \pi | {\tilde \phi} | 0 \ckt  }{q^2} \sim \const \;.  \label{WTI2bis}  
\ee
The constant  $F_{\pi}$  represents the amplitude for the broken current to produce the pion from the vacuum  (the pion decay constant).  The field ${\tilde \phi}$ is known as the pion interpolating field.

\end{document}